\documentclass[aps,pre,amsmath,twocolumn,amssymb,floatfixng,showpacs,longbibliography,
superscriptaddress,nofootinbib]{revtex4-2}
\usepackage{amsfonts}
\usepackage{amsmath}
\usepackage{amssymb}
\usepackage{latexsym}
\usepackage{array}
\newcolumntype{P}[1]{>{\centering\arraybackslash}p{#1}}
\newcolumntype{M}[1]{>{\centering\arraybackslash}m{#1}}
\usepackage{caption}
\usepackage[colorlinks,citecolor=red,urlcolor=blue,bookmarks=false,hypertexnames=true]{hyperref} 
\usepackage{xcolor}
\usepackage{bm}
\usepackage{relsize}
\usepackage{float}
\usepackage{subfloat}
\usepackage{dcolumn}
\usepackage{epsfig}
\usepackage{graphicx}
\usepackage{multirow}
\usepackage{makecell}
\usepackage{mathtools}
\usepackage[bbgreekl]{mathbbol}

\usepackage{footnote}

\begin{document}

\title{Some aspects of Microscopic Mechanism of Superconductivity in conventional and non-conventional systems}  
\author{Ranjan Chaudhury}
\altaffiliation[]{ranjan@bose.res.in}
\author{Koushik Mandal}%
\email{koushikmandal@bose.res.in}
\affiliation{%
 S N Bose National Centre for Basic Sciences \\
 Block- JD; Sector-III, Salt Lake, Kolkata, India\\
}%

\begin{abstract}
A broad review of theoretical research work involving different types of  microscopic mechanism in various classes of superconductors, carried out in our research group over a decade or so, is presented. These mechanisms include both conventional as well as exotic ones. Special emphasis is placed on the possible applications to the experimental situations. Moreover, comparison of our works with various theoretical proposals made by various other researchers, with regard to high temperature superconductivity in particular, is made. The crucial importance and special significance of our results are highlighted.
 
\end{abstract}

                               
\maketitle

\section{Introduction}
 The field of superconductivity has seen many many exciting developments and new challenges over the last 100 years, ever since its discovery in 1911. Starting with metals and metallic alloys in the early years, superconducting materials now include organic, ceramic and various other types of inorganic systems like transition metal- rare earth compounds as well as hydrides and super-hydrides. Beside the discoveries of the enriching interplay and coexistence of superconductivity with many other phenomena such as magnetism, charge density wave, structural transition in various novel systems, there has been a steady attempt to attain higher superconducting transition temperatures. The final dream has of course been to achieve room temperature superconductivity. The families belonging to the A-15 compounds, Cuprates, Fullerenes, Fe-based compounds and the most recent ones of Hydrides and super-Hydrides are all gifts in this goal oriented journey.\\
Since the discovery of the Cuprate superconductors, the research in superconductivity in general and in the domain of high temperature superconductivity in particular, received a major boost. Along with the experimental breakthroughs and advancements, the theoretical research was initiated with new vigour too. The theoretical approaches can be broadly categorized as :- (i) continuing along the conventional lines of Fermionic pairing theory and (ii) novel approaches involving both non-conventional Fermionic pairing as well as Bose-Einstein condensation.\\ 
Our small research group at S.N. Bose Centre (under the leadership of RC) has been involved with theoretical investigation of superconductivity in general and high temperature superconductivity in particular, for the last 3 decades. We will report and discuss here our work done during the last 10-15 years.\\
Our work related to the Cuprate and Pnictide based high temperature superconductors (HTSC) is mostly based on the treatment of Coulomb correlated systems (often of low dimensionality) in the normal phase and exploring the possibility of superconducting pairing thereafter. Again, the relative strength of Coulomb correlation can vary, leading to different parental normal phase scenarios and thereby affects the superconducting pairing processes and mechanisms as well. These open up the possibilities of some of the exotic mechanisms too. \\
Our work on other types of superconductors consists of :- (i) exploring boson-exchange mechanism of pairing involving phonons as well non-phonons (like Charge-transfer excitons) for both 3-dimensional and low-dimensional systems and (ii) investigation of possible occurrences of Kohn singularity and Kohn anomaly in the superconducting phases of certain elemental superconductors and its consequences for the pairing mechanism.\\
Broadly speaking, in our group we have worked on models and mechanisms corresponding to different classes of superconductors. They can be categorised into two groups :-
(A) Quasi-1D, Quasi-2D (layered) and 3D superconducting systems based on Fermionic pairing mechanism through Boson exchange,
and 
(B) Quasi-2D doped quantum Heisenberg antiferromagnets exhibiting superconductivity through conventional as well as non-conventional routes. The conventional route corresponds to determination of effective interaction between charge carriers (holes) through generalized charge stiffness constant and looking for the conditions under which this interaction can turn attractive. The non-conventional route hinges upon the existence of mobile topological magnetic excitations  and their possible pairing after getting electrically charged on doping.\\
Let us now discuss our work on both these classes of superconductors systematically in some details.
(1) The superconductors of fully 3-dimensional type as well as of lower dimensions with quasi-1D and quasi-2D characters have fascinated condensed matter physicists for many many decades. This has stimulated a lot of investigation and research into the origin of microscopic mechanisms for Cooper pairing as well as superconductivity in these diverse systems.\\
In our group, we have specifically focused on calculations relevant to (i) quasi-1D organic superconductors like $(TMTSF)_2 X$, $(TTF)TCNQ$, etc., (ii) quasi-2D superconductors belonging to the Cuprate and Pnictide families and (iii) pure 3D superconductors of elemental type like Pb and Nb as well as representative of other metallic alloys like the A-15 compounds and the Hydrides and super-Hydrides.\\
In contrast to the 3D systems, the low-dimensional superconductors are governed by the two distinct processes viz. intra-chain/ intra-layer Cooper pair formation accompanied by inter-chain/ inter-layer pair tunneling \cite{RoychowdhuryandChaudhury2015}. We had analysed both Cooper pair formations as well as superconducting properties in all these different types of systems viz. low-dimensional as well as $3$D. Furthermore, we had made use of several types of microscopic descriptions for various  normal phase scenarios,  in these calculations. Besides, we had handled both lattice and continuum models.

\section{Theoretical Methodology and Important Results}
\subsection{Boson Exchange Superconductors}
\subsubsection{\textbf{Quasi-1D Superconductors:Bechhgard salt}}
Let us first take the case of quasi-1D superconductors with a lattice model.\\
The Cooper's one pair (CP) equation for the binding energy with the filled Fermi Sea background, takes the following form \cite{RoychowdhuryandChaudhury2015} :-
\begin{equation}
\left[-\frac{\hbar^{2}}{2m}\left(\nabla_{1}^{2}+ \nabla_{2}^{2}\right) + V(\Vec{r}_{1},\Vec{r}_{2})\right] \phi(\Vec{r}_{1}-\Vec{r}_{2}) = E \phi(\Vec{r}_{1}-\Vec{r}_{2})      
\end{equation}
where, $\phi(\Vec{r}_{1}-\Vec{r}_{2})$ is the spin-singlet pair wave unction in the relative coordinate space, given by 
\begin{equation}
    \phi(\Vec{r}_{1}-\Vec{r}_{2}) = \sum_{\Vec{k}} a_{\Vec{k}} e^{i\Vec{k}.(\Vec{r}_{1}-\Vec{r}_{2})}
\end{equation}
In our present case of pairing for zero  centre of mass momenta(CMM) corresponding to the 1D nearest neighbour tight binding lattice system, the above equation(Eq.(1)) takes the form
\begin{equation}
    2\epsilon_{k} a_{\Vec{k}} - \sum_{\Vec{k^{'}}} a_{\Vec{k^{'}}} \frac{u}{L} = E a_{\Vec{k}}
\end{equation}
where $\epsilon_{k} =\epsilon_{0} -2tcos(ka)$ ; $(E-2E_{F})$ is the Cooper pair binding energy. \\
Similar equation has been set up for pair with finite CMM on the lattice\cite{RoychowdhuryandChaudhury2015}.
Here, we considered the attractive interaction mediated by the charge transfer excitons which were observed experimentally in the optical absorption experiments \cite{SchwartzDresselandGruner1998}. Besides, the magnitudes of the band filling factor as well as that of the forbidden band gap play very important roles in this analysis.\\
The major highlights of our results are :-\\
($1$)  Very successful explanation of the occurrence of superconductivity in Bechhgard salts,  a class of organic superconductors of quasi-1D type.\\
($2$) The distinct possibility of a crossover from the momentum-space pairing to a real space-like pairing,  with  increase in the magnitude of the attractive interaction.\\
The superconducting gaps are estimated from the CP equation. These are consistent with other types of experiments \cite{RoychowdhuryandChaudhury2015}.\\
Furthermore, our analysis brings out the realistic nature of Fermi Liquid-like (FL) description of the parental normal phase in contrast to that provided by the Luttinger-Tomonaga liquid (LTL) theory at low temperatures for these systems.\\ 
\subsubsection{\textbf{Quasi-2D and Multi-layered Superconductors}}
$\textbf{(a).}$
For pure 2D systems, both lattice and continuum versions were considered. In the continuum case, both filled Fermi Sea (vacuum) background and the active Fermi Sea background were taken into account. In the lattice situation, again both filling factor (or occupancy) of the band and the magnitude of the  band gap play crucial roles in the equation \cite{RoychowdhuryandChaudhury2019}.\\
The Cooper's one pair problem is relooked. The idea is to find the possibilities of the coexistence of the usual Cooper pairs (formed as a quasi-bound Fermion pair) with truly bound Fermionic pairs in the continuum situation. In other words, the aim of the investigation is  to examine the formation and simultaneous presence of both momentum-space and real-space  particle-particle pairs. In the lattice case, the challenge is to determine the criteria for the CP formation itself.  In this context, it is also quite interesting to draw attention to our work on 2D electron gas interacting via bare Coulombic (repulsive) interaction of logarithmic type \cite{ChaudhuryandGangopadhyay1995}.\\
(i) The CP equation for binding energy with zero CMM in the continuum,  with passive Fermi sea background takes the form \cite{ChaudhuryandChakraverty2000,Pal2017} \\
 \begin{equation}
     \Tilde{\epsilon} = \frac{2\hbar \omega_{c}}{\left[exp\left(\frac{4\pi \hbar^{2}}{Um}\right)-1\right]}
 \end{equation}
(ii) The CP equation for binding energy with zero CMM in the continuum with active Fermi sea background becomes \cite{Pal2017}
\begin{equation}
 \Tilde{\epsilon} = \frac{2\hbar \omega_{c}}{\left[exp\left(\frac{4\pi \hbar^{2}}{Um}\right)-1\right]^{\frac{1}{2}}}    
\end{equation}
It has a true bound state solution for very large magnitude of attractive interaction $U$.\\
(iii) The CP equation for binding energy in the lattice case  is  only valid when several constraints involving the bandwidth and the exciton energy are satisfied\cite{RoychowdhuryandChaudhury2015,RoychowdhuryandChaudhury2019}. Besides, formation of pairs is possible upto a certain  degree of filling only. Here, both Fermi liquid-like and Marginal Fermi liquid-like normal phase scenarios were considered\cite{RoychowdhuryandChaudhury2019}. \\
The salient features of our calculational results are the following :-\\
(i) In the continuum case, CPs themselves become truly bound in the situation with the passive Fermi sea background for  zero CMM pairs with arbitrarily small attractive interaction. However, for finite CMM pairs,  a large  threshold value of attractive interaction is required.  \\
(ii) In the case with  active Fermi sea background though,  the truly bound pairs always need a finite threshold for the strength of the attractive interaction for formation. Here interestingly,  the magnitude of the threshold decreases with the increase in the CMM.\\
Thus for the passive Fermi sea background $i.e.$ conventional situation, the usual Cooper pairs become truly bound spontaneously and this leads to short coherence lengths in a 2D layer for a layered superconductor. Furthermore, the pairs tend to behave more as ``real-space pairs" in this case. \\
For the active Fermi sea background however, the above scenario can hold only for Cooper pairs with CMM of the order of $\hbar k_{F}$ and that too with a finite magnitude of attractive interaction. Detailed  analytical and numerical exercises done by us, show that this threshold magnitude can be quite substantial and may even exceed the expected strength of the attractive interaction available in the pair forming 2D layers in the systems of interest \cite{Pal2017,ChaudhuryandGangopadhyay1995}. Therefore, proposed scenarios like (i) the normal phase itself consisting of a mixture of bound-Fermionic pairs and the usual single Fermions as well as (ii) the superconducting phase exhibiting both momentum-space and real-space pairs, need more detailed investigation \cite{Pal2017,ChaudhuryandGangopadhyay1995}. \\
The above consequences emanating from our calculational results, throw lot of light and doubts on the claims made earlier through several theoretical proposals for the layered superconductors \cite{Randeria1989,Randeria1989-2,Randeria1990}.\\\\
\textbf{(b).}
Widening the scope and possibilities further in a more realistic direction, we considered the situation corresponding to multi-layered (and multi-band) anisotropic superconductors more explicitly. As discussed earlier briefly, there are two distinct processes contributing here viz. intra-layer pairing combined with inter-layer pair hopping \cite{RoychowdhuryandChaudhury2015,Chaudhury2009}. This leads to the emergence of two kinds of superconducting gap parameters viz. (i) in-plane (intra-layer) gap and (ii) out-of-plane gap. They are defined below. They both are functions of the Fermionic wave-vectors. We carried out calculations with the conventional Bardeen-Cooper-Schrieffer(BCS)-type of isotropic singlet in-plane pairing through Boson mediated attraction \cite{Chaudhury2009,ChaudhuryDas2021}.\\
Making use of the generalized Hamiltonian 
\begin{eqnarray}
  H_{gen} = \sum_{m,k} \epsilon_{k,m} c_{k\sigma,m}^{+}c_{k\sigma,m} + \sum_{k,k^{'},m} V_{kk^{'},m} b_{k^{'},m}^{+}b_{k,m} \nonumber\\
 + \sum_{<mn>,k,k^{'}} \lambda_{k,k^{'}}^{mn} \left[ b_{k^{'},m}^{+}b_{k,n} +h.c.\right]
\end{eqnarray}
where, the first and the second term in the above Hamiltonian together represent the standard BCS Hamiltonian corresponding to the inter-layer pairing with $m$ being the layer index for a particular pair-forming layer. The $c's$ are usual single fermion operators and $b's$ are the Cooper pair operators. The last term in the above Hamiltonian represents the hopping of Cooper pairs between two successive layers. It may be recalled that the Cooper pair operators are related to the fermion operators in the following way
\begin{equation}
    b_{k} = c_{-k-\sigma}c_{k\sigma}
\end{equation}
The basic idea here is that the Cooper pairs are formed in each of the above mentioned layers through an effective attractive interaction $V_{kk^{'},m}$; however since superconductivity is a 3-dimensional phenomenon, we will have to invoke an inter-layer process coupling the pairs from different layers with a parameter $\lambda$.\\
The above calculation was done with the assumption of only the  nearest neighbour layers taking part in the inter-layer pair hopping process. The mean field treatment of the BCS-like Hamiltonian for the multi-layered systems leads to very interesting consequences. The most important one amongst these is the existence of a minimum magnitude of the out-of-plane gap. This threshold value however, becomes vanishingly small when the in-plane (intra-layer) component of the wave-vector is arbitrarily close to the  Fermi circle. Thus, in this case the out-of-plane gap exhibits nodes. However, for wave-vectors away from the Fermi circle this threshold value takes a finite magnitude. These interesting theoretical results await experimental verification. Nevertheless, (i) the presence of a power law behavior in the temperature dependence of the specific heat and (ii) the occurrence of a bump of the specific heat vs. temperature curve at a finite temperature, both the features having been observed experimentally in the superconducting phases of some of the multi-layered (two-band) superconductors, like $Fe-AS$, Cuprates etc are consistent with our theoretical results \cite{ChaudhuryandDas2021}.\\
\subsubsection{\textbf{3D Correlated Superconductors}}
For pure 3-dimensional superconductors, our interest was focused on handling the effect of the repulsive Coulomb correlation in the presence of the boson mediated attractive pairing interaction. We examined two distinct cases viz. (i) presence of Coulomb correlation only in the superconducting phase and (ii) presence of Coulomb correlation in the parental normal phase itself, before superconducting pairing takes place \cite{MandalandChaudhury2021,MandalandChaudhury2021arxiv}.\\
The theoretical methodology consists of the implementation of  a combination of the BCS pairing ansatz and the Gutzwiller form of projecting out the doubly occupied sites, within a variational scheme. The two proposed variational many body wave functions are stated as 
\begin{eqnarray}
  \Psi_{C} = \prod P_{G} \otimes P_{BCS} \vert FS\rangle\\
  \Psi_{CBCS} = \prod P_{BCS} \otimes P_{G} \vert FS\rangle
\end{eqnarray}
where $\vert FS\rangle$ is the non-interacting Fermi sea ground state and $P_{BCS}$ is the usual BCS pairing operator. Moreover, $P_{G}$ is the Gutzwiller partial projection operator that reduces the spatial dimension of the Slater determinant by excluding the doubly occupied sites. This conceptual idea has been implemented in the above mention two cases in order to define a correlated metallic ground state for superconducting pairing as well as to implement the Coulomb correlation on the superconducting paired state\cite{MandalandChaudhury2021,MandalandChaudhury2021arxiv}.\\
\begin{figure}
\centering
\includegraphics[scale=0.65]{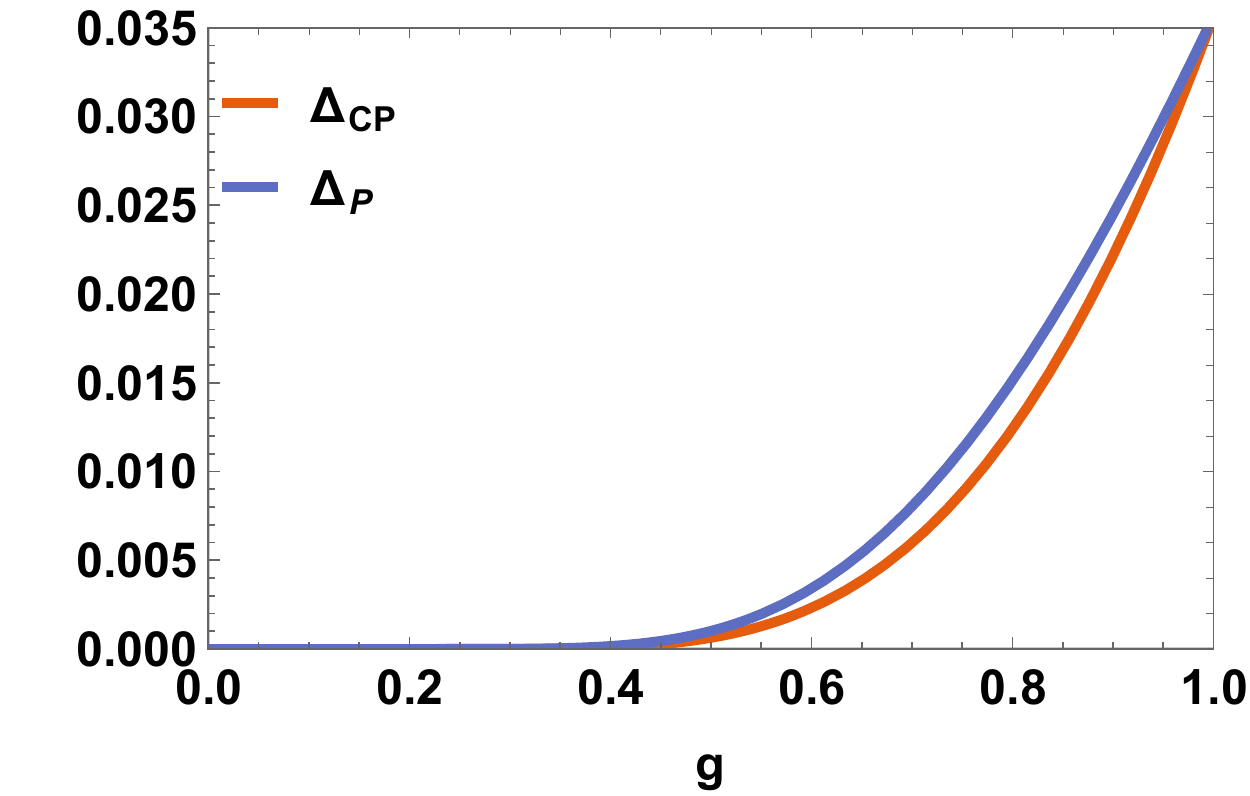}
  \caption{Superconducting pairing gap in the presence of passive($\Delta_{p}$) and active Coulomb($\Delta_{CP}$) correlation are shown here against the reduced Gutzwiller parameter $g(=1-\alpha)$ \cite{MandalandChaudhury2021}.}
  \label{fig:boat1}
\end{figure}
When subjected to the above calculational methodology, the two cases mentioned above exhibit very different consequences quite interestingly, because of the non-commuting nature of the BCS pairing operator and the Gutzwiller projection operator.  In  both the cases, the superconducting gap is generally suppressed with the increase in the magnitude of the Coulomb correlation parameter($\alpha$)(see fig:1 and 2). However, in the situation corresponding to (ii) above, a ``two-gap-like structure" emerges for the superconducting gap function with drastically different behavior for the two `gap components'. While the 1st gap decreases monotonically with the increase in the strength of the Coulomb correlation parameter, the 2nd gap shows a peaking behavior as a function of correlation parameter(see Fig:2).\\
\begin{figure}
\centering
\includegraphics[scale=0.60]{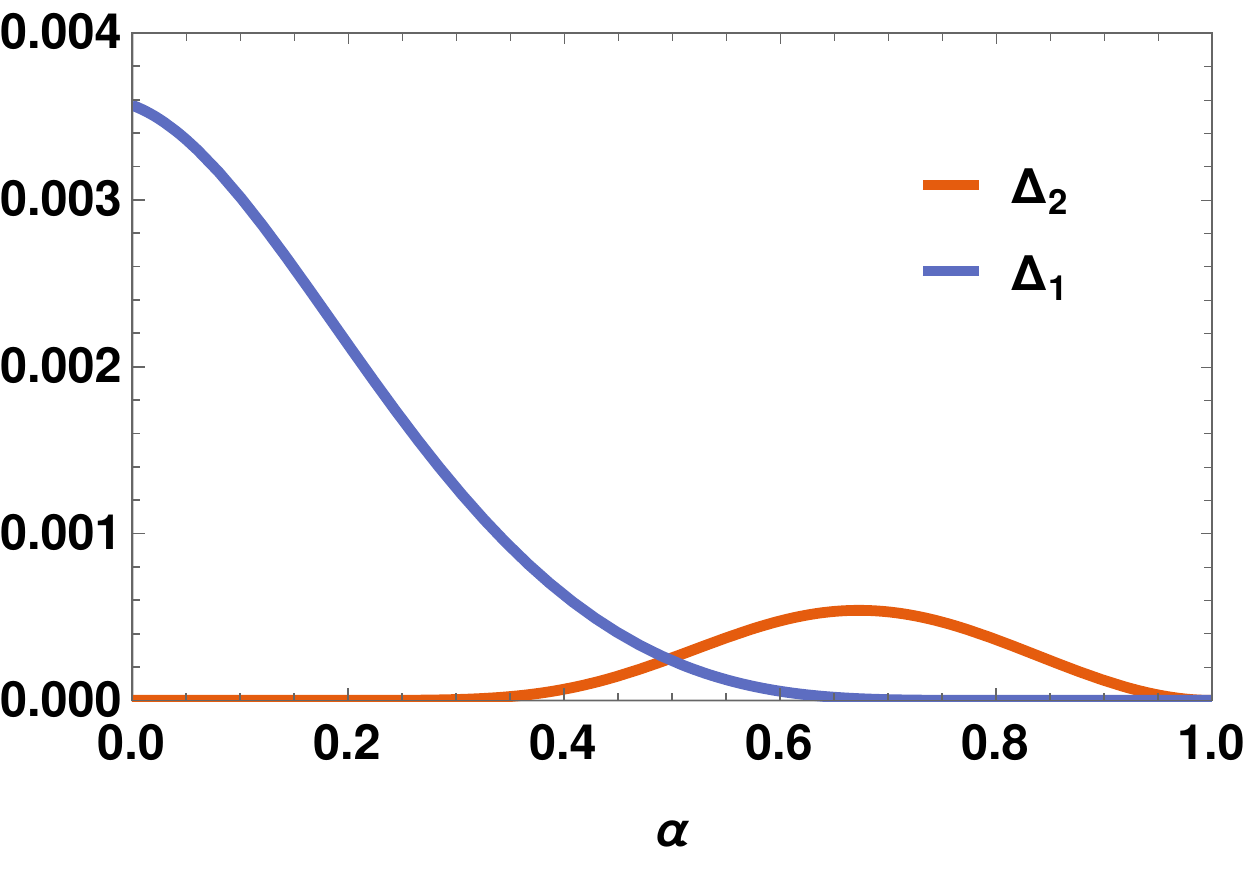}
  \caption{Variation of two gaps $\Delta_{1}$ and $\Delta_{2}$ with the Gutzwiller parameter($\alpha$) is shown for a hopping parameter $t=0.25 eV$ and in the weak coupling regime \cite{MandalandChaudhury2021arxiv}.}
  \label{fig:boat1}
\end{figure}
\subsubsection{\textbf{3D Elemental Superconductors}}
There was a further work of ours on elemental superconductors of 3D type with phonon mediated pairing like in $Pb$ and $Nb$. Our theoretical investigation motivated by the experimental work of Aynajian and coworkers using neutron spin-echo technique, brought out the phenomena of Kohn singularity and Kohn anomaly in the superconducting phases \cite{ChaudhuryandDas2013,Aynajianetal2008}. It is a striking parallelism with the two phenomena observed in the normal metals ! Our work shows that (i) in the vicinity of the Fermi wave-vector, the k-space derivative of the electronic polarizability function displays a singularity when the superconducting coherence length is very large or very very small in comparison to the lattice parameter ; and (ii) at the special point where the phonon energy equals twice the superconducting gap value, the phonon dispersion curve exhibits a point of inflexion indicating an `anomaly'. Furthermore, assuming the acoustic phonons undergoing this anomaly to be  the mediator of the pairing interaction as well, we derived an expression for the superconducting critical temperature \cite{ChaudhuryandDas2013}.\\
A purely isotropic BCS gap function ($\Delta_{0}$) at $T = 0 K$
satisfies the following approximate equation under
such a condition (assuming a ``jellium" model and keeping in
mind that at the Kohn anomaly $\hbar \omega(q) = 2\Delta_{0}$):
\begin{equation}
    \Delta_{0} \sim 3.5\Delta_{0} exp\left[-\frac{h^{2}(3\pi^{2})^{2/3}}{3n^{1/3}m\vert g_{eff}\vert^{2}4\pi^{2}}\right]
\end{equation}
with
\begin{subequations}
\begin{align}
\frac{2\Delta_{0}}{k_{B}T_{C}} = 3.5\\
T_{C} = 1.13 \Theta_{C} exp\left(-\frac{1}{g}\right)
\end{align}
\end{subequations}
Here, $T_{c}$ is the superconducting transition temperature and
$\Theta_{C}$ is the temperature equivalent of the characteristic energy of
the phonon modes mediating the attractive pairing interaction. 
We further incorporate the constraint
that the renormalized phonon frequency of the branch
undergoing the Kohn anomaly is the characteristic cutoff for
the regime of pairing interaction around the Fermi surface, i.e.
\begin{equation}
    \Theta_{C} \sim \frac{2\Delta_{0}}{k_{B}}
\end{equation}
Our work is expected to be very very useful in analysing the superconducting properties of the phonon driven elemental superconductors.\\
\subsection{Quasi-2D Doped Quantum Heisenberg Antiferromagnet}
\subsubsection{\textbf{Generalized Stiffness Calculations}}
The large class of Cuprate superconductors are hole doped Heisenberg anti-ferromagnets\cite{Endohetal1988,Yamadaetal1989}. The nature of the effective interaction between the itinerant carries in the $Cu-O$ layers in this doped phase, decides the superconducting pairing possibility. Keeping this in mind, we had theoretically calculated the generalized charge stiffness constant for the $t-J$ like models on 2D square lattice \cite{BhattacharjeeandChaudhury2021}. We scanned the entire doping regime from under-doped to over-doped. We could relate the generalized charge stiffness constant with the longitudinal dielectric function for itinerant systems by making use of the standard equations from electromagnetic theory and linear response theory. This leads to:-
\begin{equation}
    V_{eff}(\omega) = \frac{V_{0}}{1-\frac{4\pi D_{c}}{\omega^{2}}}
\end{equation}
with $V_{0}$ being the bare Coulomb interaction; $V_{eff}$ is the dynamic effective interaction and $D_{C}$ is the charge stiffness constant. The above equation shows the possibility of $V_{eff}$ turning attractive for $\omega \to 0$, in the RPA-like treatment of correlated phase\cite{BhattacharjeeandChaudhury2021}. Our results bring out a clear possibility of the existence of attractive interaction between the mobile carries in the $k-$space in the optimally doped and over-doped regimes only. This strongly supports a superconducting pairing scenario through electronic mechanism in the above region, although it is difficult to pinpoint any particular type of boson responsible for mediating the pairing interaction at this stage.\\
\subsubsection{\textbf{Field-theory and Phenomenology} }
We employed field-theoretic techniques and analysed results from inelastic neutron scattering experiments using semi-classical phenomenology, to identify topological excitations of `meronic' type in the insulating anti-ferromagnetic phase of the Cuprates in a certain temperature range \cite{Chaudhury1992, SarkarChaudhuryandPaul2017}. These merons (and anti-merons) are generalizations of vortices (and anti-vortices) as appearing in the Berezinskii-Kosterlitz-Thouless (BKT) theory \cite{Berezinskii1971,KosterlitzandThouless1972}. Our detailed calculations on both quantum and semi-classical versions of XY-anisotropic realistic anti-ferromagnetic spin model firmly establish the existence of these topological excitations in the medium wave-length regime\cite{ChaudhuryPaul2013,ChaudhuryPaul2009}. This is further supported by the appearance of a `Central peak' in the dynamical structure function, as observed in the inelastic neutron scattering experiments on the Cuprates \cite{Endohetal1988,Yamadaetal1989}. Again, there have been several suggestions and proposals related to the possibilities of these kinds of topological excitations becoming electrically charged and pairing up as well, in the lightly doped (under-doped) phase of the Cuprates \cite{Chenetal1989,Lykkenetal1991,Girvin1992}. Thus, our investigation is a right step in this direction of verifying these exotic proposals and  can definitely throw new light on the quantum nature of these electrically charged topological excitations and their role in bringing about superconducting pairing.\\
\section{Concluding Remarks}
We have presented  a concise  overview of the work carried out in our research group on superconductivity, based on diverse theoretical approaches. They cover systems from conventional to the novel ones. To keep it accessible to a large class of readers,  the presentation has been made semi-qualitative and semi-quantitative. Nevertheless, we have attempted to bring out the very important contributions of ours towards understanding the microscopic aspects of superconductivity in very challenging situations.
 \section{Acknowledgement}
 RC would like to thank all his collaborators and his research students for their valuable contributions in the works and projects reported here. He would also like to acknowledge all the infra-structural facilities provided by S. N. Bose Centre during these research activities in our group. KM would like to thank UGC, India for the financial support during his PhD tenure.  
\bibliography{reference.bib} 

\begin{thebibliography}{27}%
\makeatletter
\providecommand \@ifxundefined [1]{%
 \@ifx{#1\undefined}
}%
\providecommand \@ifnum [1]{%
 \ifnum #1\expandafter \@firstoftwo
 \else \expandafter \@secondoftwo
 \fi
}%
\providecommand \@ifx [1]{%
 \ifx #1\expandafter \@firstoftwo
 \else \expandafter \@secondoftwo
 \fi
}%
\providecommand \natexlab [1]{#1}%
\providecommand \enquote  [1]{``#1''}%
\providecommand \bibnamefont  [1]{#1}%
\providecommand \bibfnamefont [1]{#1}%
\providecommand \citenamefont [1]{#1}%
\providecommand \href@noop [0]{\@secondoftwo}%
\providecommand \href [0]{\begingroup \@sanitize@url \@href}%
\providecommand \@href[1]{\@@startlink{#1}\@@href}%
\providecommand \@@href[1]{\endgroup#1\@@endlink}%
\providecommand \@sanitize@url [0]{\catcode `\\12\catcode `\$12\catcode
  `\&12\catcode `\#12\catcode `\^12\catcode `\_12\catcode `\%12\relax}%
\providecommand \@@startlink[1]{}%
\providecommand \@@endlink[0]{}%
\providecommand \url  [0]{\begingroup\@sanitize@url \@url }%
\providecommand \@url [1]{\endgroup\@href {#1}{\urlprefix }}%
\providecommand \urlprefix  [0]{URL }%
\providecommand \Eprint [0]{\href }%
\providecommand \doibase [0]{https://doi.org/}%
\providecommand \selectlanguage [0]{\@gobble}%
\providecommand \bibinfo  [0]{\@secondoftwo}%
\providecommand \bibfield  [0]{\@secondoftwo}%
\providecommand \translation [1]{[#1]}%
\providecommand \BibitemOpen [0]{}%
\providecommand \bibitemStop [0]{}%
\providecommand \bibitemNoStop [0]{.\EOS\space}%
\providecommand \EOS [0]{\spacefactor3000\relax}%
\providecommand \BibitemShut  [1]{\csname bibitem#1\endcsname}%
\let\auto@bib@innerbib\@empty
\bibitem [{\citenamefont {Chowdhury}\ and\ \citenamefont
  {Chaudhury}(2015)}]{RoychowdhuryandChaudhury2015}%
  \BibitemOpen
  \bibfield  {author} {\bibinfo {author} {\bibfnamefont {S.~R.}\ \bibnamefont
  {Chowdhury}}\ and\ \bibinfo {author} {\bibfnamefont {R.}~\bibnamefont
  {Chaudhury}},\ }\bibfield  {title} {\bibinfo {title} {Investigation of
  fermionic pairing on tight binding lattice for low dimensional
  systems-\text{Fermi liquid vs. Luttinger-Tomonaga liquid}},\ }\href@noop {}
  {\bibfield  {journal} {\bibinfo  {journal} {Physica B: Condensed Matter}\
  }\textbf {\bibinfo {volume} {465}},\ \bibinfo {pages} {60} (\bibinfo {year}
  {2015})}\BibitemShut {NoStop}%
\bibitem [{\citenamefont {Schwartz}\ \emph {et~al.}(1998)\citenamefont
  {Schwartz}, \citenamefont {Dressel}, \citenamefont {Gr{\"u}ner},
  \citenamefont {Vescoli}, \citenamefont {Degiorgi},\ and\ \citenamefont
  {Giamarchi}}]{SchwartzDresselandGruner1998}%
  \BibitemOpen
  \bibfield  {author} {\bibinfo {author} {\bibfnamefont {A.}~\bibnamefont
  {Schwartz}}, \bibinfo {author} {\bibfnamefont {M.}~\bibnamefont {Dressel}},
  \bibinfo {author} {\bibfnamefont {G.}~\bibnamefont {Gr{\"u}ner}}, \bibinfo
  {author} {\bibfnamefont {V.}~\bibnamefont {Vescoli}}, \bibinfo {author}
  {\bibfnamefont {L.}~\bibnamefont {Degiorgi}},\ and\ \bibinfo {author}
  {\bibfnamefont {T.}~\bibnamefont {Giamarchi}},\ }\bibfield  {title} {\bibinfo
  {title} {On-chain electrodynamics of metallic ${(TMTSF)_{2}X}$ salts:
  Observation of \text{Tomonaga-Luttinger }liquid response},\ }\href@noop {}
  {\bibfield  {journal} {\bibinfo  {journal} {Phys. Rev. B}\ }\textbf {\bibinfo
  {volume} {58}},\ \bibinfo {pages} {1261} (\bibinfo {year}
  {1998})}\BibitemShut {NoStop}%
\bibitem [{\citenamefont {Chowdhury}\ and\ \citenamefont
  {Chaudhury}(2019)}]{RoychowdhuryandChaudhury2019}%
  \BibitemOpen
  \bibfield  {author} {\bibinfo {author} {\bibfnamefont {S.~R.}\ \bibnamefont
  {Chowdhury}}\ and\ \bibinfo {author} {\bibfnamefont {R.}~\bibnamefont
  {Chaudhury}},\ }\bibfield  {title} {\bibinfo {title} {Theoretical
  investigation of the feasibility of electronic mechanism for superconducting
  pairing in over doped cuprates},\ }\href@noop {} {\bibfield  {journal}
  {\bibinfo  {journal} {Journal of Low Temperature Physics}\ }\textbf {\bibinfo
  {volume} {196}},\ \bibinfo {pages} {335} (\bibinfo {year}
  {2019})}\BibitemShut {NoStop}%
\bibitem [{\citenamefont {Chaudhury}\ and\ \citenamefont
  {Gangopadhyay}(1995)}]{ChaudhuryandGangopadhyay1995}%
  \BibitemOpen
  \bibfield  {author} {\bibinfo {author} {\bibfnamefont {R.}~\bibnamefont
  {Chaudhury}}\ and\ \bibinfo {author} {\bibfnamefont {D.}~\bibnamefont
  {Gangopadhyay}},\ }\bibfield  {title} {\bibinfo {title} {Interacting
  fermions, scaling and possible departure from \text{Fermi} liquid behavior},\
  }\href@noop {} {\bibfield  {journal} {\bibinfo  {journal} {Mod. Phys. Lett.
  B}\ }\textbf {\bibinfo {volume} {9}},\ \bibinfo {pages} {1657} (\bibinfo
  {year} {1995})}\BibitemShut {NoStop}%
\bibitem [{\citenamefont {Chaudhury}\ and\ \citenamefont
  {Chakraverty}(2000)}]{ChaudhuryandChakraverty2000}%
  \BibitemOpen
  \bibfield  {author} {\bibinfo {author} {\bibfnamefont {R.}~\bibnamefont
  {Chaudhury}}\ and\ \bibinfo {author} {\bibfnamefont {B.~K.}\ \bibnamefont
  {Chakraverty}},\ }\bibfield  {title} {\bibinfo {title} {Criterion for bound
  state formation in layered systems},\ }\href@noop {} {\bibfield  {journal}
  {\bibinfo  {journal} {SNBS-CNRS Research Report under Indo-French
  Collaborative Project on High-Temperature Superconductivity}\ } (\bibinfo
  {year} {2000})}\BibitemShut {NoStop}%
\bibitem [{\citenamefont {Pal}(2017)}]{Pal2017}%
  \BibitemOpen
  \bibfield  {author} {\bibinfo {author} {\bibfnamefont {S.}~\bibnamefont
  {Pal}},\ }\bibfield  {title} {\bibinfo {title} {Pairing in two dimensions and
  possible consequences for superconductivity: extension of \text{Cooper's}
  approach},\ }\href@noop {} {\bibfield  {journal} {\bibinfo  {journal}
  {Journal of Physics Communications}\ }\textbf {\bibinfo {volume} {1}},\
  \bibinfo {pages} {055029} (\bibinfo {year} {2017})}\BibitemShut {NoStop}%
\bibitem [{\citenamefont {Randeria}\ \emph
  {et~al.}(1989{\natexlab{a}})\citenamefont {Randeria}, \citenamefont {Duan},\
  and\ \citenamefont {Shieh}}]{Randeria1989}%
  \BibitemOpen
  \bibfield  {author} {\bibinfo {author} {\bibfnamefont {M.}~\bibnamefont
  {Randeria}}, \bibinfo {author} {\bibfnamefont {J.-M.}\ \bibnamefont {Duan}},\
  and\ \bibinfo {author} {\bibfnamefont {L.-Y.}\ \bibnamefont {Shieh}},\
  }\bibfield  {title} {\bibinfo {title} {Bound states, \text{Cooper} pairing,
  and \text{Bose} condensation in two dimensions},\ }\href@noop {} {\bibfield
  {journal} {\bibinfo  {journal} {Physical review letters}\ }\textbf {\bibinfo
  {volume} {62}},\ \bibinfo {pages} {981} (\bibinfo {year}
  {1989}{\natexlab{a}})}\BibitemShut {NoStop}%
\bibitem [{\citenamefont {Randeria}\ \emph
  {et~al.}(1989{\natexlab{b}})\citenamefont {Randeria}, \citenamefont {Duan},\
  and\ \citenamefont {Shieh}}]{Randeria1989-2}%
  \BibitemOpen
  \bibfield  {author} {\bibinfo {author} {\bibfnamefont {M.}~\bibnamefont
  {Randeria}}, \bibinfo {author} {\bibfnamefont {J.-M.}\ \bibnamefont {Duan}},\
  and\ \bibinfo {author} {\bibfnamefont {L.-Y.}\ \bibnamefont {Shieh}},\
  }\bibfield  {title} {\bibinfo {title} {Bound states, \text{Cooper} pairing,
  and \text{Bose} condensation in two dimensions},\ }\href@noop {} {\bibfield
  {journal} {\bibinfo  {journal} {Physical review letters}\ }\textbf {\bibinfo
  {volume} {62}},\ \bibinfo {pages} {2887} (\bibinfo {year}
  {1989}{\natexlab{b}})}\BibitemShut {NoStop}%
\bibitem [{\citenamefont {Randeria}\ \emph {et~al.}(1990)\citenamefont
  {Randeria}, \citenamefont {Duan},\ and\ \citenamefont
  {Shieh}}]{Randeria1990}%
  \BibitemOpen
  \bibfield  {author} {\bibinfo {author} {\bibfnamefont {M.}~\bibnamefont
  {Randeria}}, \bibinfo {author} {\bibfnamefont {J.-M.}\ \bibnamefont {Duan}},\
  and\ \bibinfo {author} {\bibfnamefont {L.-Y.}\ \bibnamefont {Shieh}},\
  }\bibfield  {title} {\bibinfo {title} {Superconductivity in a two-dimensional
  \text{Fermi gas: Evolution from Cooper pairing to Bose} condensation},\
  }\href@noop {} {\bibfield  {journal} {\bibinfo  {journal} {Physical Review
  B}\ }\textbf {\bibinfo {volume} {41}},\ \bibinfo {pages} {327} (\bibinfo
  {year} {1990})}\BibitemShut {NoStop}%
\bibitem [{\citenamefont {Chaudhury}(2009)}]{Chaudhury2009}%
  \BibitemOpen
  \bibfield  {author} {\bibinfo {author} {\bibfnamefont {R.}~\bibnamefont
  {Chaudhury}},\ }\bibfield  {title} {\bibinfo {title} {A possible theoretical
  model for studying superconductivity in \text{Fe}-based systems},\
  }\href@noop {} {\bibfield  {journal} {\bibinfo  {journal} {arXiv preprint
  arXiv:0901.1438v1[cond-mat.supr-con]}\ } (\bibinfo {year}
  {2009})}\BibitemShut {NoStop}%
\bibitem [{\citenamefont {Chaudhury}\ and\ \citenamefont
  {Das}(2022)}]{ChaudhuryDas2021}%
  \BibitemOpen
  \bibfield  {author} {\bibinfo {author} {\bibfnamefont {R.}~\bibnamefont
  {Chaudhury}}\ and\ \bibinfo {author} {\bibfnamefont {M.~P.}\ \bibnamefont
  {Das}},\ }\bibfield  {title} {\bibinfo {title} {Superconducting pairing in
  multi-layered systems within single band approximation and possible extension
  to multi-band situation},\ }\href@noop {} {\bibfield  {journal} {\bibinfo
  {journal} {(To be Communicated)}\ } (\bibinfo {year} {2022})}\BibitemShut
  {NoStop}%
\bibitem [{\citenamefont {Mandal}\ and\ \citenamefont
  {Chaudhury}(2021{\natexlab{a}})}]{MandalandChaudhury2021}%
  \BibitemOpen
  \bibfield  {author} {\bibinfo {author} {\bibfnamefont {K.}~\bibnamefont
  {Mandal}}\ and\ \bibinfo {author} {\bibfnamefont {R.}~\bibnamefont
  {Chaudhury}},\ }\bibfield  {title} {\bibinfo {title} {Interplay of pairing
  correlation and \text{Coulomb} correlation in boson exchange
  superconductors},\ }\href@noop {} {\bibfield  {journal} {\bibinfo  {journal}
  {The European Physical Journal B}\ }\textbf {\bibinfo {volume} {94}},\
  \bibinfo {pages} {1} (\bibinfo {year} {2021}{\natexlab{a}})}\BibitemShut
  {NoStop}%
\bibitem [{\citenamefont {Mandal}\ and\ \citenamefont
  {Chaudhury}(2021{\natexlab{b}})}]{MandalandChaudhury2021arxiv}%
  \BibitemOpen
  \bibfield  {author} {\bibinfo {author} {\bibfnamefont {K.}~\bibnamefont
  {Mandal}}\ and\ \bibinfo {author} {\bibfnamefont {R.}~\bibnamefont
  {Chaudhury}},\ }\bibfield  {title} {\bibinfo {title} {A theoretical analysis
  of superconducting pairing in correlated metallic systems},\ }\href@noop {}
  {\bibfield  {journal} {\bibinfo  {journal} {arXiv preprint
  arXiv:2111.10888v1[cond-mat.supr-con]}\ } (\bibinfo {year}
  {2021}{\natexlab{b}})}\BibitemShut {NoStop}%
\bibitem [{\citenamefont {Chaudhury}\ and\ \citenamefont
  {Das}(2013)}]{ChaudhuryandDas2013}%
  \BibitemOpen
  \bibfield  {author} {\bibinfo {author} {\bibfnamefont {R.}~\bibnamefont
  {Chaudhury}}\ and\ \bibinfo {author} {\bibfnamefont {M.~P.}\ \bibnamefont
  {Das}},\ }\bibfield  {title} {\bibinfo {title} {Kohn singularity and
  \text{Kohn} anomaly in conventional superconductors—role of pairing
  mechanism},\ }\href@noop {} {\bibfield  {journal} {\bibinfo  {journal}
  {Journal of Physics: Condensed Matter}\ }\textbf {\bibinfo {volume} {25}},\
  \bibinfo {pages} {122202} (\bibinfo {year} {2013})}\BibitemShut {NoStop}%
\bibitem [{\citenamefont {Aynajian}\ \emph {et~al.}(2008)\citenamefont
  {Aynajian}, \citenamefont {Keller}, \citenamefont {Boeri}, \citenamefont
  {Shapiro}, \citenamefont {Habicht},\ and\ \citenamefont
  {Keimer}}]{Aynajianetal2008}%
  \BibitemOpen
  \bibfield  {author} {\bibinfo {author} {\bibfnamefont {P.}~\bibnamefont
  {Aynajian}}, \bibinfo {author} {\bibfnamefont {T.}~\bibnamefont {Keller}},
  \bibinfo {author} {\bibfnamefont {L.}~\bibnamefont {Boeri}}, \bibinfo
  {author} {\bibfnamefont {S.}~\bibnamefont {Shapiro}}, \bibinfo {author}
  {\bibfnamefont {K.}~\bibnamefont {Habicht}},\ and\ \bibinfo {author}
  {\bibfnamefont {B.}~\bibnamefont {Keimer}},\ }\bibfield  {title} {\bibinfo
  {title} {Energy gaps and \text{Kohn} anomalies in elemental
  superconductors},\ }\href@noop {} {\bibfield  {journal} {\bibinfo  {journal}
  {Science}\ }\textbf {\bibinfo {volume} {319}},\ \bibinfo {pages} {1509}
  (\bibinfo {year} {2008})}\BibitemShut {NoStop}%
\bibitem [{\citenamefont {Endoh}\ \emph {et~al.}(1988)\citenamefont {Endoh},
  \citenamefont {Yamada}, \citenamefont {Birgeneau}, \citenamefont {Gabbe},
  \citenamefont {Jenssen}, \citenamefont {Kastner}, \citenamefont {Peters},
  \citenamefont {Picone}, \citenamefont {Thurston}, \citenamefont {Tranquada},
  \citenamefont {Shirane}, \citenamefont {Hakida}, \citenamefont {Oda},
  \citenamefont {Enomoto}, \citenamefont {Suzuki},\ and\ \citenamefont
  {Murakami}}]{Endohetal1988}%
  \BibitemOpen
  \bibfield  {author} {\bibinfo {author} {\bibfnamefont {Y.}~\bibnamefont
  {Endoh}}, \bibinfo {author} {\bibfnamefont {K.}~\bibnamefont {Yamada}},
  \bibinfo {author} {\bibfnamefont {R.~J.}\ \bibnamefont {Birgeneau}}, \bibinfo
  {author} {\bibfnamefont {D.~R.}\ \bibnamefont {Gabbe}}, \bibinfo {author}
  {\bibfnamefont {H.~P.}\ \bibnamefont {Jenssen}}, \bibinfo {author}
  {\bibfnamefont {M.~A.}\ \bibnamefont {Kastner}}, \bibinfo {author}
  {\bibfnamefont {C.~J.}\ \bibnamefont {Peters}}, \bibinfo {author}
  {\bibfnamefont {P.~J.}\ \bibnamefont {Picone}}, \bibinfo {author}
  {\bibfnamefont {T.~R.}\ \bibnamefont {Thurston}}, \bibinfo {author}
  {\bibfnamefont {J.~M.}\ \bibnamefont {Tranquada}}, \bibinfo {author}
  {\bibfnamefont {G.}~\bibnamefont {Shirane}}, \bibinfo {author} {\bibfnamefont
  {Y.}~\bibnamefont {Hakida}}, \bibinfo {author} {\bibfnamefont
  {M.}~\bibnamefont {Oda}}, \bibinfo {author} {\bibfnamefont {Y.}~\bibnamefont
  {Enomoto}}, \bibinfo {author} {\bibfnamefont {M.}~\bibnamefont {Suzuki}},\
  and\ \bibinfo {author} {\bibfnamefont {T.}~\bibnamefont {Murakami}},\
  }\bibfield  {title} {\bibinfo {title} {Static and dynamic spin correlations
  in pure and doped ${La_{2}CuO_{4}}$},\ }\href@noop {} {\bibfield  {journal}
  {\bibinfo  {journal} {Physical Review B}\ }\textbf {\bibinfo {volume} {37}},\
  \bibinfo {pages} {7443} (\bibinfo {year} {1988})}\BibitemShut {NoStop}%
\bibitem [{\citenamefont {Yamada}\ \emph {et~al.}(1989)\citenamefont {Yamada},
  \citenamefont {Kakurai}, \citenamefont {Endoh}, \citenamefont {Thurston},
  \citenamefont {Kastner}, \citenamefont {Birgeneau}, \citenamefont {Shirane},
  \citenamefont {Hidaka},\ and\ \citenamefont {Murakami}}]{Yamadaetal1989}%
  \BibitemOpen
  \bibfield  {author} {\bibinfo {author} {\bibfnamefont {K.}~\bibnamefont
  {Yamada}}, \bibinfo {author} {\bibfnamefont {K.}~\bibnamefont {Kakurai}},
  \bibinfo {author} {\bibfnamefont {Y.}~\bibnamefont {Endoh}}, \bibinfo
  {author} {\bibfnamefont {T.}~\bibnamefont {Thurston}}, \bibinfo {author}
  {\bibfnamefont {M.}~\bibnamefont {Kastner}}, \bibinfo {author} {\bibfnamefont
  {R.}~\bibnamefont {Birgeneau}}, \bibinfo {author} {\bibfnamefont
  {G.}~\bibnamefont {Shirane}}, \bibinfo {author} {\bibfnamefont
  {Y.}~\bibnamefont {Hidaka}},\ and\ \bibinfo {author} {\bibfnamefont
  {T.}~\bibnamefont {Murakami}},\ }\bibfield  {title} {\bibinfo {title} {Spin
  dynamics in the two-dimensional quantum antiferromagnet ${La_{2}CuO_{4}}$},\
  }\href@noop {} {\bibfield  {journal} {\bibinfo  {journal} {Physical Review
  B}\ }\textbf {\bibinfo {volume} {40}},\ \bibinfo {pages} {4557} (\bibinfo
  {year} {1989})}\BibitemShut {NoStop}%
\bibitem [{\citenamefont {Bhattacharjee}\ and\ \citenamefont
  {Chaudhury}(2021)}]{BhattacharjeeandChaudhury2021}%
  \BibitemOpen
  \bibfield  {author} {\bibinfo {author} {\bibfnamefont {S.}~\bibnamefont
  {Bhattacharjee}}\ and\ \bibinfo {author} {\bibfnamefont {R.}~\bibnamefont
  {Chaudhury}},\ }\bibfield  {title} {\bibinfo {title} {Study of effective
  coupling between charge degrees of freedom in low dimensional hole-doped
  quantum antiferromagnets},\ }\href@noop {} {\bibfield  {journal} {\bibinfo
  {journal} {Canadian Journal of Physics}\ }\textbf {\bibinfo {volume} {99}},\
  \bibinfo {pages} {159} (\bibinfo {year} {2021})}\BibitemShut {NoStop}%
\bibitem [{\citenamefont {Chaudhury}(1992)}]{Chaudhury1992}%
  \BibitemOpen
  \bibfield  {author} {\bibinfo {author} {\bibfnamefont {R.}~\bibnamefont
  {Chaudhury}},\ }\bibfield  {title} {\bibinfo {title} {High temperature
  superconductivity-current status, our theoretical and experimental work},\
  }\href@noop {} {\bibfield  {journal} {\bibinfo  {journal} {Indian Journal of
  Physics}\ }\textbf {\bibinfo {volume} {66}},\ \bibinfo {pages} {159}
  (\bibinfo {year} {1992})}\BibitemShut {NoStop}%
\bibitem [{\citenamefont {Sarkar}\ \emph {et~al.}(2017)\citenamefont {Sarkar},
  \citenamefont {Chaudhury},\ and\ \citenamefont
  {Paul}}]{SarkarChaudhuryandPaul2017}%
  \BibitemOpen
  \bibfield  {author} {\bibinfo {author} {\bibfnamefont {S.}~\bibnamefont
  {Sarkar}}, \bibinfo {author} {\bibfnamefont {R.}~\bibnamefont {Chaudhury}},\
  and\ \bibinfo {author} {\bibfnamefont {S.~K.}\ \bibnamefont {Paul}},\
  }\bibfield  {title} {\bibinfo {title} {Semi-phenomenological analysis of
  neutron scattering results for quasi-two dimensional quantum
  anti-ferromagnet},\ }\href@noop {} {\bibfield  {journal} {\bibinfo  {journal}
  {Journal of Magnetism and Magnetic Materials}\ }\textbf {\bibinfo {volume}
  {421}},\ \bibinfo {pages} {207} (\bibinfo {year} {2017})}\BibitemShut
  {NoStop}%
\bibitem [{\citenamefont {Berezinskii}(1972)}]{Berezinskii1971}%
  \BibitemOpen
  \bibfield  {author} {\bibinfo {author} {\bibfnamefont {V.~L.}\ \bibnamefont
  {Berezinskii}},\ }\bibfield  {title} {\bibinfo {title} {Destruction of
  long-range order in one-dimensional and two-dimensional systems possessing a
  continuous symmetry group. \text{II.} quantum systems},\ }\href@noop {}
  {\bibfield  {journal} {\bibinfo  {journal} {Sov. Phys. JETP}\ }\textbf
  {\bibinfo {volume} {34}},\ \bibinfo {pages} {610} (\bibinfo {year}
  {1972})}\BibitemShut {NoStop}%
\bibitem [{\citenamefont {Kosterlitz}\ and\ \citenamefont
  {Thouless}(1972)}]{KosterlitzandThouless1972}%
  \BibitemOpen
  \bibfield  {author} {\bibinfo {author} {\bibfnamefont {J.~M.}\ \bibnamefont
  {Kosterlitz}}\ and\ \bibinfo {author} {\bibfnamefont {D.}~\bibnamefont
  {Thouless}},\ }\bibfield  {title} {\bibinfo {title} {Long range order and
  metastability in two dimensional solids and superfluids.(application of
  dislocation theory)},\ }\href@noop {} {\bibfield  {journal} {\bibinfo
  {journal} {Journal of Physics C: Solid State Physics}\ }\textbf {\bibinfo
  {volume} {5}},\ \bibinfo {pages} {L124} (\bibinfo {year} {1972})}\BibitemShut
  {NoStop}%
\bibitem [{\citenamefont {Chaudhury}\ and\ \citenamefont
  {Paul}(2013)}]{ChaudhuryPaul2013}%
  \BibitemOpen
  \bibfield  {author} {\bibinfo {author} {\bibfnamefont {R.}~\bibnamefont
  {Chaudhury}}\ and\ \bibinfo {author} {\bibfnamefont {S.~K.}\ \bibnamefont
  {Paul}},\ }\bibfield  {title} {\bibinfo {title} {Topological excitations in
  quantum spin systems},\ }\href@noop {} {\bibfield  {journal} {\bibinfo
  {journal} {Advances in Condensed Matter Physics}\ }\textbf {\bibinfo {volume}
  {2013}} (\bibinfo {year} {2013})}\BibitemShut {NoStop}%
\bibitem [{\citenamefont {Chaudhury}\ and\ \citenamefont
  {Paul}(2009)}]{ChaudhuryPaul2009}%
  \BibitemOpen
  \bibfield  {author} {\bibinfo {author} {\bibfnamefont {R.}~\bibnamefont
  {Chaudhury}}\ and\ \bibinfo {author} {\bibfnamefont {S.~K.}\ \bibnamefont
  {Paul}},\ }\bibfield  {title} {\bibinfo {title} {Physical realization and
  possible identification of topological excitations in quantum
  \text{Heisenberg} anti-ferromagnet on a two dimensional lattice},\
  }\href@noop {} {\bibfield  {journal} {\bibinfo  {journal} {The European
  Physical Journal B}\ }\textbf {\bibinfo {volume} {69}},\ \bibinfo {pages}
  {491} (\bibinfo {year} {2009})}\BibitemShut {NoStop}%
\bibitem [{\citenamefont {Chen}\ \emph {et~al.}(1989)\citenamefont {Chen},
  \citenamefont {Wilczek}, \citenamefont {Witten},\ and\ \citenamefont
  {Halperin}}]{Chenetal1989}%
  \BibitemOpen
  \bibfield  {author} {\bibinfo {author} {\bibfnamefont {Y.-H.}\ \bibnamefont
  {Chen}}, \bibinfo {author} {\bibfnamefont {F.}~\bibnamefont {Wilczek}},
  \bibinfo {author} {\bibfnamefont {E.}~\bibnamefont {Witten}},\ and\ \bibinfo
  {author} {\bibfnamefont {B.~I.}\ \bibnamefont {Halperin}},\ }\bibfield
  {title} {\bibinfo {title} {On anyon superconductivity},\ }\href@noop {}
  {\bibfield  {journal} {\bibinfo  {journal} {International Journal of Modern
  Physics B}\ }\textbf {\bibinfo {volume} {3}},\ \bibinfo {pages} {1001}
  (\bibinfo {year} {1989})}\BibitemShut {NoStop}%
\bibitem [{\citenamefont {Lykken}\ \emph {et~al.}(1991)\citenamefont {Lykken},
  \citenamefont {Sonnenschein},\ and\ \citenamefont {Weiss}}]{Lykkenetal1991}%
  \BibitemOpen
  \bibfield  {author} {\bibinfo {author} {\bibfnamefont {J.~D.}\ \bibnamefont
  {Lykken}}, \bibinfo {author} {\bibfnamefont {J.}~\bibnamefont
  {Sonnenschein}},\ and\ \bibinfo {author} {\bibfnamefont {N.}~\bibnamefont
  {Weiss}},\ }\bibfield  {title} {\bibinfo {title} {The theory of anyonic
  superconductivity: A review},\ }\href@noop {} {\bibfield  {journal} {\bibinfo
   {journal} {International Journal of Modern Physics A}\ }\textbf {\bibinfo
  {volume} {6}},\ \bibinfo {pages} {5155} (\bibinfo {year} {1991})}\BibitemShut
  {NoStop}%
\bibitem [{\citenamefont {Girvin}\ \emph {et~al.}(1992)\citenamefont {Girvin},
  \citenamefont {Wallin}, \citenamefont {Cha}, \citenamefont {Fisher},\ and\
  \citenamefont {Young}}]{Girvin1992}%
  \BibitemOpen
  \bibfield  {author} {\bibinfo {author} {\bibfnamefont {S.}~\bibnamefont
  {Girvin}}, \bibinfo {author} {\bibfnamefont {M.}~\bibnamefont {Wallin}},
  \bibinfo {author} {\bibfnamefont {M.-C.}\ \bibnamefont {Cha}}, \bibinfo
  {author} {\bibfnamefont {M.}~\bibnamefont {Fisher}},\ and\ \bibinfo {author}
  {\bibfnamefont {A.~P.}\ \bibnamefont {Young}},\ }\bibfield  {title} {\bibinfo
  {title} {Universal conductivity at the superconductor-insulator transition in
  two-dimensions},\ }\href@noop {} {\bibfield  {journal} {\bibinfo  {journal}
  {Progress of Theoretical Physics Supplement}\ }\textbf {\bibinfo {volume}
  {107}},\ \bibinfo {pages} {135} (\bibinfo {year} {1992})}\BibitemShut
  {NoStop}%
\end{thebibliography}%
\end{document}